# Perturbative Gluon Resummation of the Top-Quark Production Cross Section

EDMOND L. BERGER AND HARRY CONTOPANAGOS

*High Energy Physics Division, Argonne National Laboratory*

*Argonne, IL 60439-4815, USA*

## ABSTRACT

We present a calculation of the total cross section for top quark production based on a new perturbative resummation of gluon radiative corrections to the basic QCD subprocesses. We use Principal Value Resummation to calculate all relevant large threshold corrections. Advantages of this method include its independence from arbitrary infrared cutoffs and specification of the perturbative regime of applicability. For $p\bar{p}$ collisions at center-of-mass energy $\sqrt{s} = 1.8\,\text{TeV}$ and a top mass of 175 GeV, we compute $\sigma(t\bar{t}) = 5.52^{+0.07}_{-0.45} pb$.

The quest for the top quark $t$ reached fruition recently with the publication of results by two collaborations of experimenters studying $t\bar{t}$ pair production in proton-antiproton collisions at the Fermilab Tevatron[1]: $p + \bar{p} \to t + \bar{t} + X$. The large mass of the quark and the possibility that its observed cross section exceeds theoretical expectations have, in turn, stimulated considerable theoretical activity. Among theoretical questions deserving attention is the quantitative reliability of cross sections based on the main production mechanism in perturbative quantum chromodynamics (pQCD), namely $t\bar{t}$ pair creation.

At lowest order (tree-level), the two partonic subprocesses are quark-antiquark annihilation:

$$q + \bar{q} \to t + \bar{t} \qquad (1)$$

and gluon-gluon fusion:

$$g + g \to t + \bar{t} \ . \qquad (2)$$

These subprocesses are of order $\alpha_s^2$ in the strong coupling strength. The calculated top quark cross sections depend on these subprocess cross sections and on the parton density distributions that specify the probability densities of the quarks, antiquarks, and gluons of the incident $p$ and $\bar{p}$.

Both order $\alpha_s^2$ [2] and the next-to-leading order $\alpha_s^3$ [3][4] contributions have been investigated thoroughly[5]. One observation is that the size of the $\mathcal{O}(\alpha_s^3)$ terms in the partonic cross sections is particularly large near the $t\bar{t}$ production threshold, raising questions about the reliability of perturbation theory. This region of phase space is important for top quark production at the Tevatron owing to the large mass of the top quark. Virtual and bremsstrahlung corrections to the tree-level production channels, Eqs. (1) and (2), dominate the next-to-leading contributions in the near threshold region. The large threshold corrections can be identified with numerically large logarithmic terms attributable to initial-state gluon bremsstrahlung, accompanying uncompensated mass-singularities removed after mass-factorization. After convolution of the partonic cross section with the parton flux, the $\mathcal{O}(\alpha_s^3)$ corrections to the physical cross section yield less dramatic enhancements, of the order of 25% at top mass $m = 175$ GeV. These enhancements motivate a complete study of the large logarithmic corrections at the partonic level, to all orders in perturbation theory. In this Letter we present a resummation of soft gluon corrections to the $t\bar{t}$ cross section employing the Principal Value Resummation (PVR) technique[6].

As in other hard-scattering processes, where large logarithmic threshold contributions are present[7], resummation of these corrections to all orders in $\alpha_s$ is important both for theoretical understanding of the perturbative process and for numerical control of the resulting predictions. One method for resummation has been implemented previously for $t\bar{t}$ pair production[8]. In most resummation methods[9], including [8], the threshold corrections are exponentiated into a function of the QCD running coupling constant, $\alpha_s$, evaluated at a variable momentum scale, and formally integrated throughout the proper phase space. The variable scale is a measure of the momenta of the gluons emitted from the initial partons, and the associated integration represents the inclusiveness of the measurement.

An inherent uncertainty and limitation of the resummation method of [8] is its dependence on an undetermined infrared (IR) cutoff. This cutoff enters when the exponentiated resummation function is integrated over phase space. In the formulation of [8], it cuts off the Landau pole of the QCD running coupling constant and the corresponding threshold region. Since the function is exponentiated, the dependence of the resummed cross section on this cutoff is important numerically. Further, dependence of the resummation exponent on this additional undetermined mass scale spoils explicit renormalization-group invariance properties for any region of the radiation phase space. Such properties can be implemented for a fixed value of the IR cutoff, but a simultaneous variation of the cutoff and the renormalization scale would again produce large sensitivities.

The main advantage of PVR is that it does not depend on arbitrary IR cutoffs, as all Landau-pole singularities are by-passed by a Cauchy principal-value prescription. We emphasize three strengths:

(i) PVR reproduces correctly all the asymptotic properties and renormalon structure of pQCD. Due to the absence of extra undetermined scales, it allows for an evaluation of the perturbative regime, i.e., the region of the radiation phase space where perturbation theory should be valid.

(ii) The approach implements universality of resummation automatically since it is the same mechanism for all relevant processes, independent of undetermined parameters. This mechanism produces a different perturbative regime for different production channels, but the mechanism itself is common to such channels.

(iii) PVR been tested successfully in $l\bar{l}$ production[10] where its definition of resummed pQCD agrees fairly well with experiment, even at relatively low values of the hard scale (in the region of 5-10 GeV).

In the remainder of this Letter, we present our results for the physical inclusive total cross section for $t\bar{t}$-production in PVR for a top-mass range $m \in \{150, 250\}$ GeV, including a discussion of the remaining theoretical uncertainties. We also show some results for the behavior of partonic cross sections at $m = 175$ GeV. We present our predictions in the $\overline{MS}$ factorization scheme, in which the $q$, $\bar{q}$ and $g$ densities and the next-to-leading order partonic cross sections are defined unambiguously. We postpone presentation of calculational details, more extensive comparisons with [8], and corresponding analyses of the differences to our longer companion paper[11].

It has been observed that the functional form of the leading threshold corrections, order-by-order in perturbation theory, appears to be universal, i.e., independent of the hard-scattering process, except for differences in multiplicative color factors and process-specific kinematics. An example is the "universality" between dilepton- ($l\bar{l}$) and $t\bar{t}$-production, emphasized in [8]. We understand this universality to arise from the fact that mass-singularities in the former process are due purely to bremsstrahlung from the initial state, given that the $l\bar{l}$ pair is produced through an electroweak vector boson of large invariant mass, an "external potential" to the QCD sector, whereas, in the latter the final quarks are (highly) massive, and gluon bremsstrahlung from (or interference with) that sector does not produce leading mass singularities and associated large logarithms. In $t\bar{t}$ production *near threshold*, the heavy quark final-state sector "factorizes" and behaves as an external heavy object, leaving the light initial states to produce the *leading large* QCD corrections. For the purposes of this work we accept this universality, assuming it valid to all orders in $\alpha_s$, as in [8]. Furthermore, we have chosen the order-by-order logarithmic structure and running coupling constant identical to those of [8].

The resummed partonic cross sections, including all large threshold corrections according to PVR, can be written as

$$\sigma_{ij}^{PV}(\eta, m^2) = \int_{1-4(1+\eta)+4\sqrt{1+\eta}}^{1} dz \left[1 + \mathcal{H}_{ij}(z, \alpha)\right] \sigma'_{ij}(\eta, m^2, z). \tag{3}$$

In Eq. (3),

$$\mathcal{H}_{ij}(z, \alpha) = \int_0^{\ln(\frac{1}{1-z})} dx\, e^{E_{ij}(x,\alpha)} \sum_{j=0}^{\infty} Q_j(x, \alpha), \tag{4}$$

$\sigma'_{ij}(\eta, m^2, z) = d(\sigma_{ij}^{(0)}(\eta, m^2, z))/dz$, and $\sigma_{ij}^{(0)}$ is the tree-level partonic cross section expressed in terms of inelastic kinematic variables [8] to account for the emitted radiation.[★] The variable $\eta = \hat{s}/4m^2 - 1$ is the distance from the partonic production threshold, $\alpha \equiv \alpha_s(m)/\pi$, and $ij \in \{q\bar{q}, gg\}$ denotes the initial parton channel. The integration in Eq. (3) is over the phase space of the radiated gluons, parametrized through the dimensionless variable $z$. In $l\bar{l}$ production $z = Q^2/\hat{s}$, where $Q$ is the invariant dilepton mass, and $z = 1$ corresponds to zero gluon momentum.

Equation (4) is the main content of PVR. The *leading* large threshold corrections are contained in the exponent $E_{ij}(x, \alpha)$, which is a calculable polynomial in $x$. $\{Q_j(x,\alpha)\}$ are calculable functions produced by the analytical inversion of the Mellin transform from moment space (where the argument of $E_{ij}$ corresponds to $x \to \ln n$) to the physically relevant momentum space expressed in Eq. (4). These functions are produced by the resummation and are expressed in terms of successive derivatives of $E^{\dagger}$: $P_k(x,\alpha) \equiv \partial^k E(x,\alpha)/k!\partial^k x$, $k = 1, 2, ....$ Given that $E$ contains at most one more power of $x$ than of $\alpha$, $P_k$ contains at least $k-1$ fewer powers of $x$ than of $\alpha$, and $Q_j$ contributes the *resummed version* of terms containing $j$ fewer powers of $x$ than of $\alpha$ in the integrand of Eq. (4).[‡] For example $Q_0$ contains all powers $P_1^m$, $Q_1$ all powers $P_2 P_1^m$, etc.

---

★ In other words, substituting $\mathcal{H} \to 0$ in Eq. (3) and using the kinematic constraint $\sigma_{ij}^0(\eta, m^2, z = 1 - 4(1+\eta) + 4\sqrt{1+\eta}) = 0$, we obtain the tree-level partonic cross section, $\sigma_{ij}^0(\eta, m^2) = \sigma_{ij}^0(\eta, m^2, z = 1)$.

The main object of the resummation, $E_{ij}$, was studied extensively in the context of $l\bar{l}$ production [10]. Embodying the universality discussed earlier, its functional form for $t\bar{t}$ production is identical to that of $l\bar{l}$ production, except for the identification of the two separate channels, denoted by the subscript $ij$. However, there are significant differences between the two processes which affect the way we implement PVR practically for the process at hand. These can be described as follows:

First, only the *leading* threshold corrections are universal. The physical reason behind this is the existence of hadronic final states, absent in $l\bar{l}$ production, that produce interference effects with the initial hadronic states, a statement that can be deduced from the $\mathcal{O}(\alpha^3)$ calculation as well. Therefore, from all structures $\{Q_j\}$ in Eq. (4), the very leading one should be considered universal. This is the linear term in $P_1$, which turns out to be $P_1$ itself. Hence, Eq. (4) can be integrated explicitly, and Eq. (3) may be written as

$$\sigma_{ij}^{PV}(\eta, m^2) = \int_{1-4(1+\eta)+4\sqrt{1+\eta}}^{1} dz \, e^{E_{ij}(\ln(\frac{1}{1-z}),\alpha)} \sigma_{ij}^{!}(\eta, m^2, z). \tag{5}$$

---

† For simplicity, we drop the channel indices.
‡ These contributions are only approximately equal to the corresponding structures obtained upon a finite-order expansion of the resummation formula, since such expansion involves truncation of a product of series.

The second issue has to do with the perturbative regime in momentum space. A basic property of $E_{ij}(\ln n, \alpha)$ in moment space is that it has a perturbative representation in the region $2b_2 \alpha \ln n < 1$, where $b_2$ is the first coefficient of the QCD beta function. In the region $1 < 2b_2 \alpha \ln n < \infty$, it has a non-perturbative representation which, in turn, is an exponential suppression. In fact, $\lim_{n \to \infty} E_{ij}(\ln n, \alpha) = -\infty$. This latter "higher-twist" region is of negligible importance, especially for the large scale $m$ of the top quark production process. These considerations apply in moment space and are clearly independent of channel-dependent color factors, relying purely on the properties of the running coupling constant and PVR itself. To characterize a region in moment space as "higher-twist", however, one must first convert to momentum space through inversion of the Mellin transform, Eq. (4). Specification of the boundary, mentioned already in ref.[6], is realized by the constraint that all $\{Q_j\}$, $j \geq 1$ be small compared to $Q_0$ which provides the leading integrand in Eq. (4), according to the previous power counting. This constraint can be shown quite generally to correspond to

$$P_1\left(\ln\left(\frac{1}{1-z}\right), \alpha\right) < 1 \ . \tag{6}$$

This perturbative regime in momentum space corresponds nicely with the perturbative regime in moment space, since $P_1$ contains at most equal powers of logarithms and $\alpha$. This determination of the perturbative regime is also less academic, in the sense that it realistically includes all relevant orders of perturbation theory, all relevant constants generated by the running of the coupling and, most importantly, the color factors that differentiate between various production channels. Color factors are not included in more abstract discussions in moment space but are physically extremely important.

One could actually apply Eq. (5) all the way to $z = 1$ by using contour integration as suggested by PVR, beyond the perturbative regime of Eq. (6), but one would then be using a *model* for non-perturbative effects, the one suggested by PVR, far beyond the knowledge justified by perturbation theory. In this work we confine our attention to the perturbative regime, and hence our cross section is evaluated accordingly, namely

$$\sigma_{ij}^{PV_{\text{pert}}}(\eta, m^2) = \int_{1-4(1+\eta)+4\sqrt{1+\eta}}^{z_{max}} dz\, e^{E_{ij}(\ln(\frac{1}{1-z}),\alpha)} \sigma'_{ij}(\eta, m^2, z) \ . \tag{7}$$

The upper limit is calculated through

$$P_1\left(\ln\left(\frac{1}{1-z_{max}}\right), \alpha\right) = 1 \ . \tag{8}$$

Equations (7), (8) are the basic formulas of our approach. Based on these, notice that our perturbative approach can probe the threshold region down to $\eta \geq (1 - z_{max})/2$. It turns out our final result does not rely much upon the PVR method to by-pass IR renormalons and associated problems, precisely because it is restricted to the perturbative regime. In that sense, the presence of arbitrary IR cutoffs in previous resummations is superfluous, as all necessary information about IR sensitivity (i.e., the perturbative regime) can be obtained by examining the perturbative asymptotic properties of the resummation functions. Further analysis of these issues, especially the perturbative representation of Eq. (4) as an asymptotic series and related numerical issues relevant to the color factors characterizing each specific production channel, will be presented in our longer paper[11].

We provide our final resummed cross sections for each production channel by taking into account the complete next-to-leading calculation, $\sigma_{ij}^{(0+1)}$, through the improved prediction

$$\sigma_{ij}^{\text{final}}(\eta, m^2) = \sigma_{ij}^{PV_{\text{pert}}}(\eta, m^2) + \sigma_{ij}^{(0+1)}(\eta, m^2) - \sigma_{ij}^{(0+1)}(\eta, m^2)\bigg|_{PV} \ . \tag{9}$$

The last term in Eq. (9) is the next-to-leading order partonic cross section included in the resummation, $\sigma_{ij}^{PV_{\text{pert}}}$.

In Fig.1 we present our prediction for the partonic cross sections in the $q\bar{q}$ and $gg$ channels, in the $\overline{MS}$ scheme, at $m = 175$ GeV. We also show the tree-level ($\sigma^{(0)}$) and next-to-leading ($\sigma^{(0+1)}$) counterparts. Notice that the three curves differ substantially in the partonic threshold region $\eta < 1$, with the final resummed curve exceeding the other two. Below $\eta \simeq 7 \times 10^{-3}$ in the $q\bar{q}$ channel and $\simeq 5 \times 10^{-2}$ in the $gg$ channel, our final perturbative cross sections become identical to the next-to-leading order cross sections. This is a consequence of our decision to perform the resummation in the perturbative domain. This domain is smaller in the $gg$ case because the color factors are larger in this channel. Above $\eta \simeq 1$, we note that our resummed cross sections are essentially identical to the next-to-leading order cross sections, as is to be expected since the near-threshold enhancements that concern us in this paper are not relevant at large $\eta$.

In Fig.2 we plot the physical cross section $\sigma^{\text{final}}$, along with the tree-level and the next-to-leading order counterparts, for the $q\bar{q}$ and $gg$ channels for a wide range of top mass. These are obtained after the partonic cross sections shown in Fig.1 are convoluted with parton densities and integrals are performed over $\eta$. We use the most recent $CTEQ3M$ parton distributions[12]. Here, the usual factorization scale $\mu$ has been chosen equal to the mass of the top. We observe that perturbative resummation enhances the cross section by about 10%.

In Fig.3 we plot the same cross section as a function of the factorization ("hard") scale $\mu$ in a range $\mu/m \in \{0.5, 2\}$. Notice the mild dependence, as well as the interesting shape that peaks around the value 1. We consider the variation of the physical cross section the range $\mu/m \in \{0.5, 2\}$ a good measure of the theoretical perturbative uncertainty.

Our prediction for the inclusive total $t\bar{t}$-production cross section at the Tevatron, using $CTEQ3M$ parton densities, is

$$\sigma^{t\bar{t}}_{\text{final}}(m = 175 \text{ GeV}) = 5.52^{+0.07}_{-0.45} \, pb \, . \tag{10}$$

The "central" value of 5.52 $pb$ is obtained with $\mu/m = 1$, and the upper and lower limits of the uncertainty band correspond to the maximum and minimum values of the cross section in the range $\mu/m \in \{0.5, 2\}$. The cross section is insensitive to the choice of parton densities. Repeating the same analysis with the $MRS(A')$ densities[13], we obtain

$$\sigma_{\text{final}}^{t\bar{t}}(m = 175 \text{ GeV}) = 5.32^{+0.08}_{-0.41} \text{ pb} . \quad (11)$$

In Tables 1 (2) we summarize the main features of the physical cross section, in pb, for top mass $m = 175$ GeV for three choices of $\mu$ and $CTEQ3M$ ($MRS(A')$) parton densities. The first line in each table shows the values of the two-loop $\alpha_s/\pi$ for the corresponding set; there is very little difference in the values. The $MRS(A')$ cross sections are a bit smaller than the $CTEQ3M$ values. For the $q\bar{q}$ channel this difference is less than 4% and is about 5% for the $gg$ channel.

| $m = 175$ GeV | $\mu = m/2$ | $\mu = m$ | $\mu = 2m$ |
|---|---|---|---|
| $\alpha_s/\pi$ | 0.03584 | 0.03260 | 0.02991 |
| $\sigma_{q\bar{q}}^{(0)}$ | 5.02 | 3.69 | 2.78 |
| $\sigma_{q\bar{q}}^{(0+1)}$ | 4.13 | 4.43 | 4.21 |
| $\sigma_{q\bar{q}}^{\text{final}}$ | 4.70 | 4.87 | 4.55 |
| $\sigma_{gg}^{(0)}$ | 0.55 | 0.34 | 0.22 |
| $\sigma_{gg}^{(0+1)}$ | 0.72 | 0.63 | 0.50 |
| $\sigma_{gg}^{\text{final}}$ | 0.75 | 0.65 | 0.52 |
| $\sigma_{\text{final}}^{t\bar{t}}$ | 5.45 | 5.52 | 5.07 |

Table 1.

| $m = 175$ GeV | $\mu = m/2$ | $\mu = m$ | $\mu = 2m$ |
|---|---|---|---|
| $\alpha_s/\pi$ | 0.03596 | 0.03269 | 0.02998 |
| $\sigma_{q\bar{q}}^{(0)}$ | 4.88 | 3.57 | 2.68 |
| $\sigma_{q\bar{q}}^{(0+1)}$ | 4.00 | 4.28 | 4.08 |
| $\sigma_{q\bar{q}}^{\text{final}}$ | 4.56 | 4.71 | 4.41 |
| $\sigma_{gg}^{(0)}$ | 0.51 | 0.32 | 0.21 |
| $\sigma_{gg}^{(0+1)}$ | 0.67 | 0.59 | 0.49 |
| $\sigma_{gg}^{\text{final}}$ | 0.70 | 0.61 | 0.50 |
| $\sigma_{\text{final}}^{t\bar{t}}$ | 5.26 | 5.32 | 4.91 |

Table 2.

The value in Eq. (10) is larger than that of [8], but within uncertainties, and it agrees with the present CDF and D0 measurements [1]. In Fig.4 we show the top-mass dependence of the physical cross section for $p\bar{p} \to (t\bar{t})X$.

The band of perturbative uncertainty quoted in Eq. (10) is relatively narrow. On the other hand, we noted in discussing Fig.1 that there is a reasonable range of $\eta$ near threshold in which perturbative resummation does not apply. Perturbation theory is not justified in this region. Correspondingly, further strong interaction enhancements of the $t\bar{t}$ cross section may arise from physics in this region. We know of no reliable way to estimate the size of such non-perturbative effects and, therefore, cannot include such uncertainties in the estimates of the *perturbative* uncertainty of Eqs. (10) and (11).

In our longer companion paper we will present mathematical and physical details, more extensive comparisons of our approach to existing results, and a discussion of remaining theoretical uncertainties.

Acknowledgements: We thank E. Laenen, J. Smith and G. Sterman for many useful discussions. This work was supported by the U.S. Department of Energy, Division of High Energy Physics, contract W-31-109-ENG-38.

FIGURE CAPTIONS

1. The total parton-parton cross sections as a function of $\eta$ in the $\overline{MS}$ scheme at $m = 175$ GeV for the subprocesses (a) $q\bar{q} \to t\bar{t}X$ and (b) $gg \to t\bar{t}X$. Plotted are the lowest order Born cross section (dotted line), the next-to-leading-order cross section (dashed line), and the cross section obtained after resummation of soft gluons by the principal value resummation method (solid line). The QCD scale $\mu = m$.

2. The calculated cross sections for $t\bar{t}$ production as a function of top mass $m$ in $p\bar{p}$ collisions at $\sqrt{s} = 1.8$ TeV for the $q\bar{q}$ (upper bunch) and $gg$ (lower bunch) subprocesses. Each bunch consists of three curves: the Born cross section (dotted line), the next-to-leading-order cross section (dashed line), and the cross section obtained after soft gluon resummation (solid line).

3. Plot showing the calculated dependence of the final resummed cross section on $(\mu/m)$ for $t\bar{t}$ production at $m = 175$ GeV and $\sqrt{s} = 1.8$ TeV. Shown also is the next-to-leading order result (dashed curve).

4. Physical cross section for $p\bar{p} \to (t\bar{t})X$ at $\sqrt{s} = 1.8$ TeV as a function of top mass. Data from the CDF and D0[1] collaborations are shown. Our calculated cross section here is obtained from adding the final resummed $q\bar{q}$ and $gg$ contributions. Shown are entries for $\mu/m = 0.5$ (dashed), 1 (solid) and 2 (dotted).

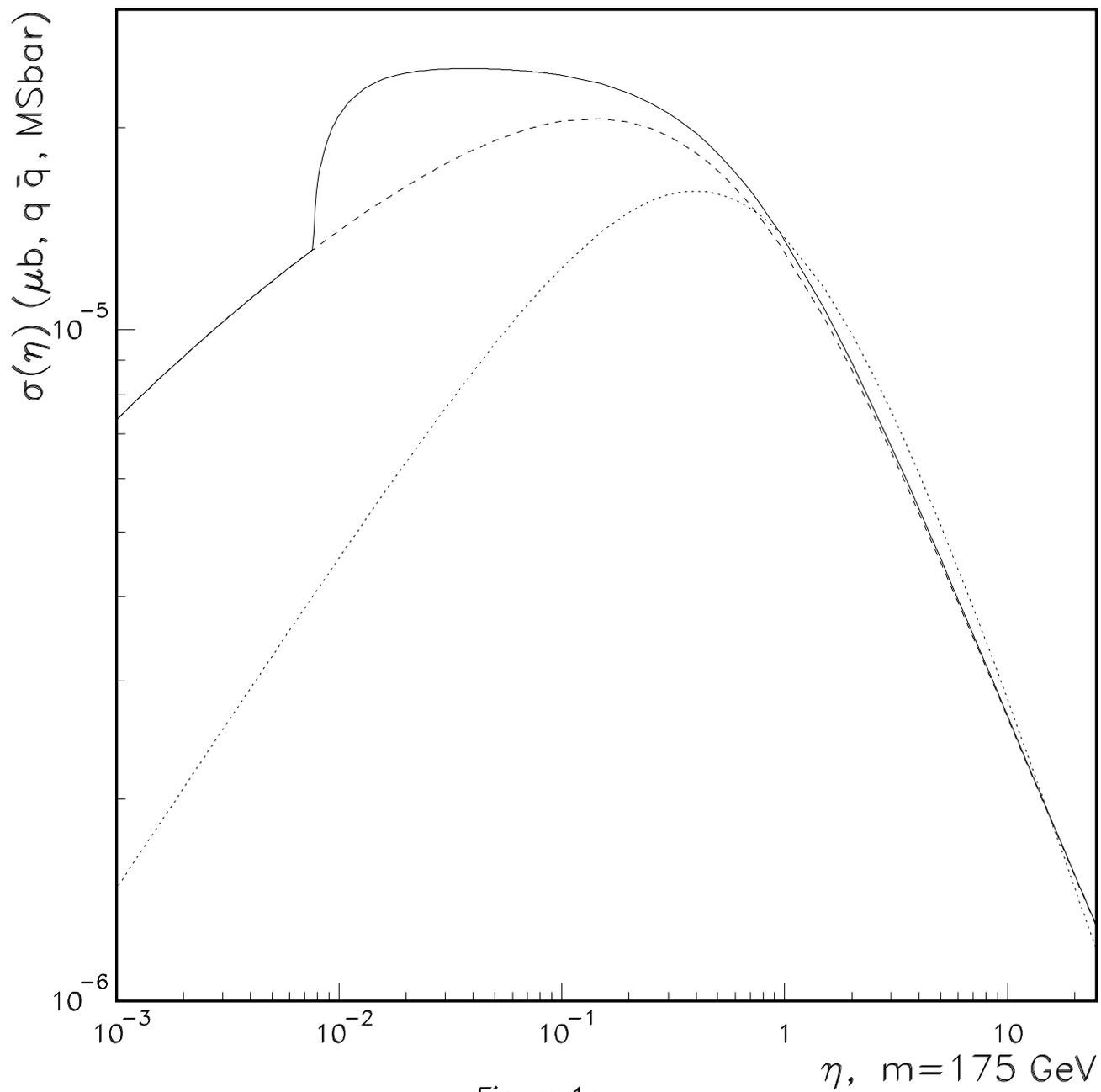

Figure 1a

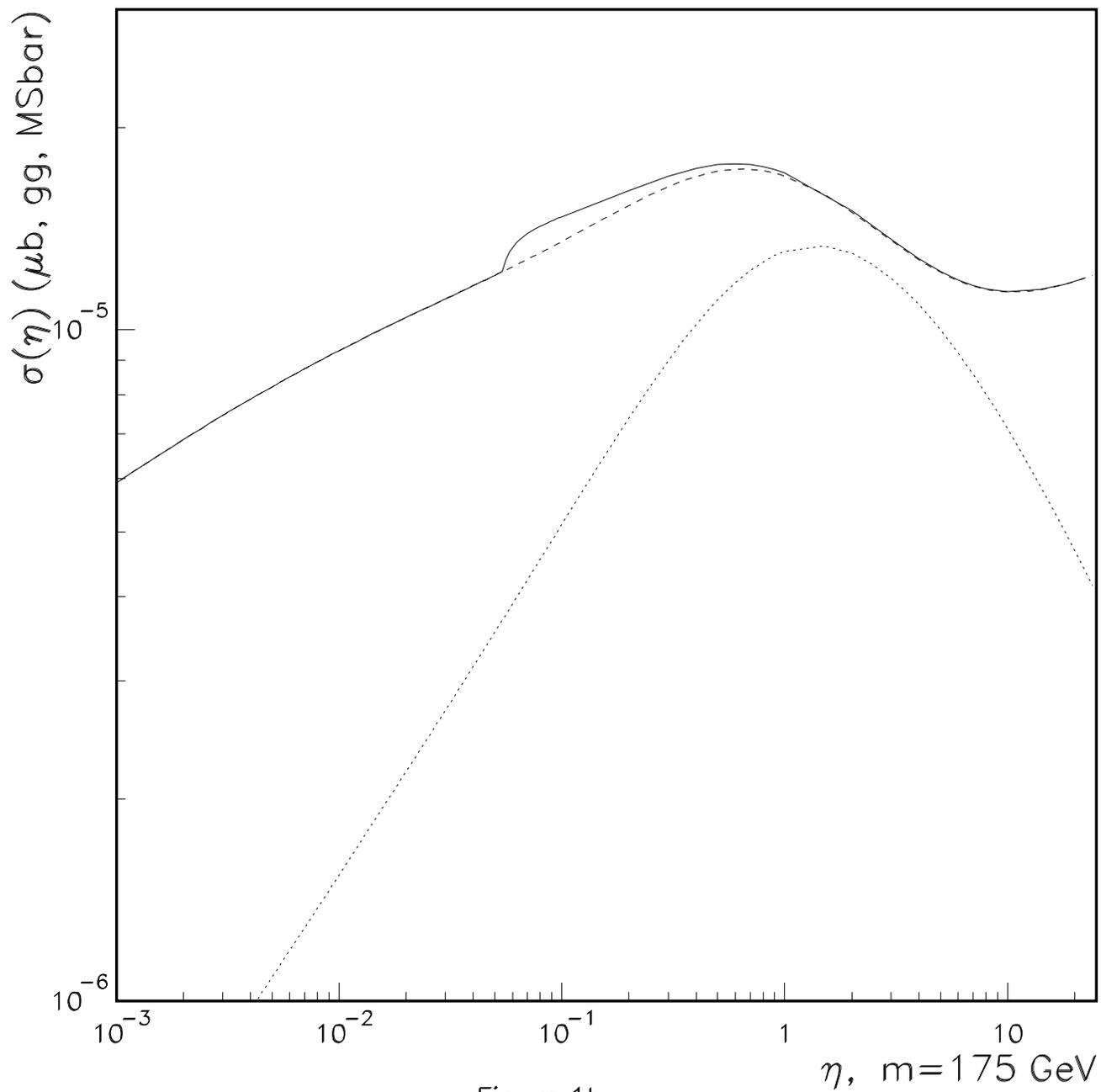

Figure 1b

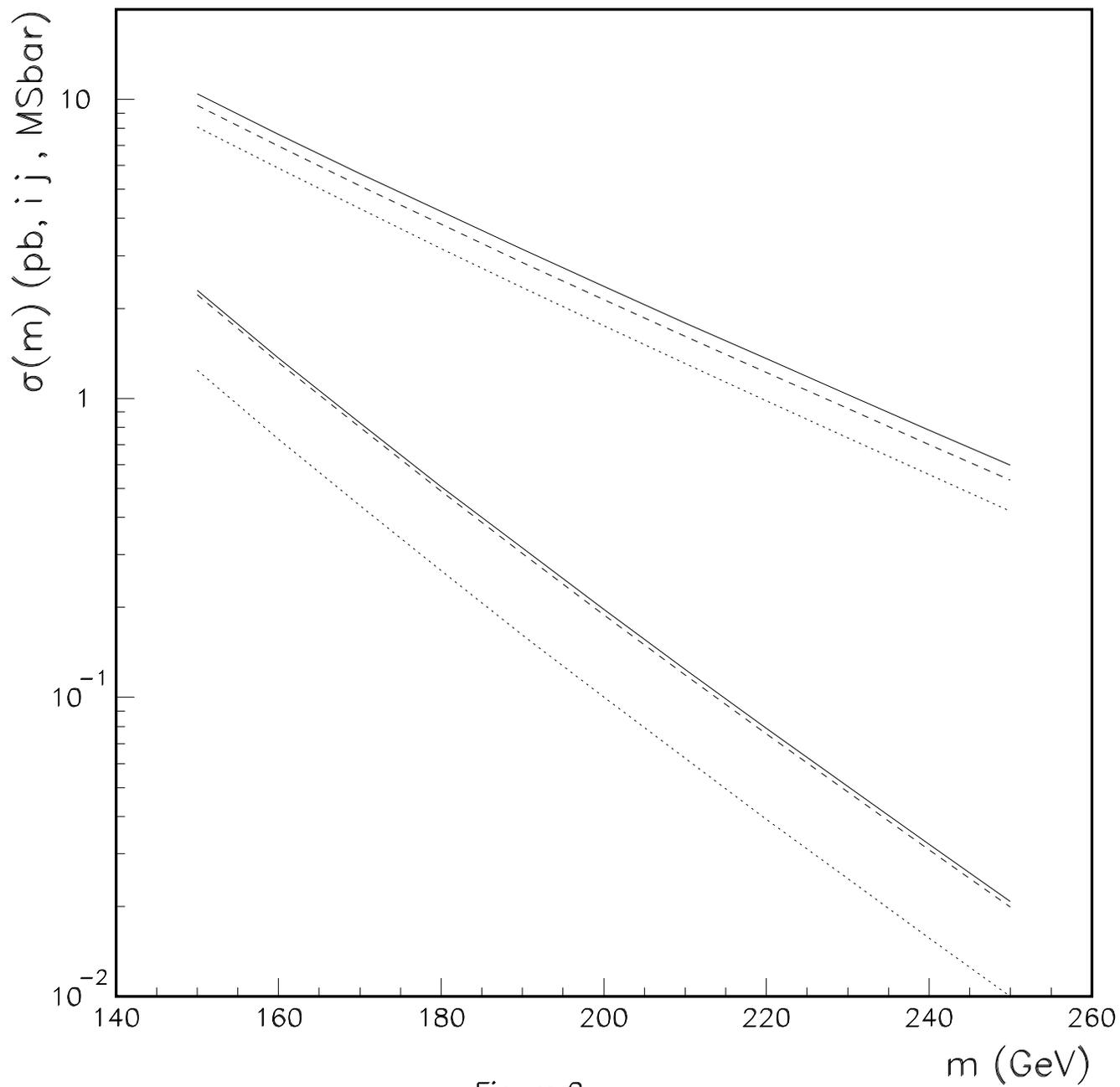

Figure 2

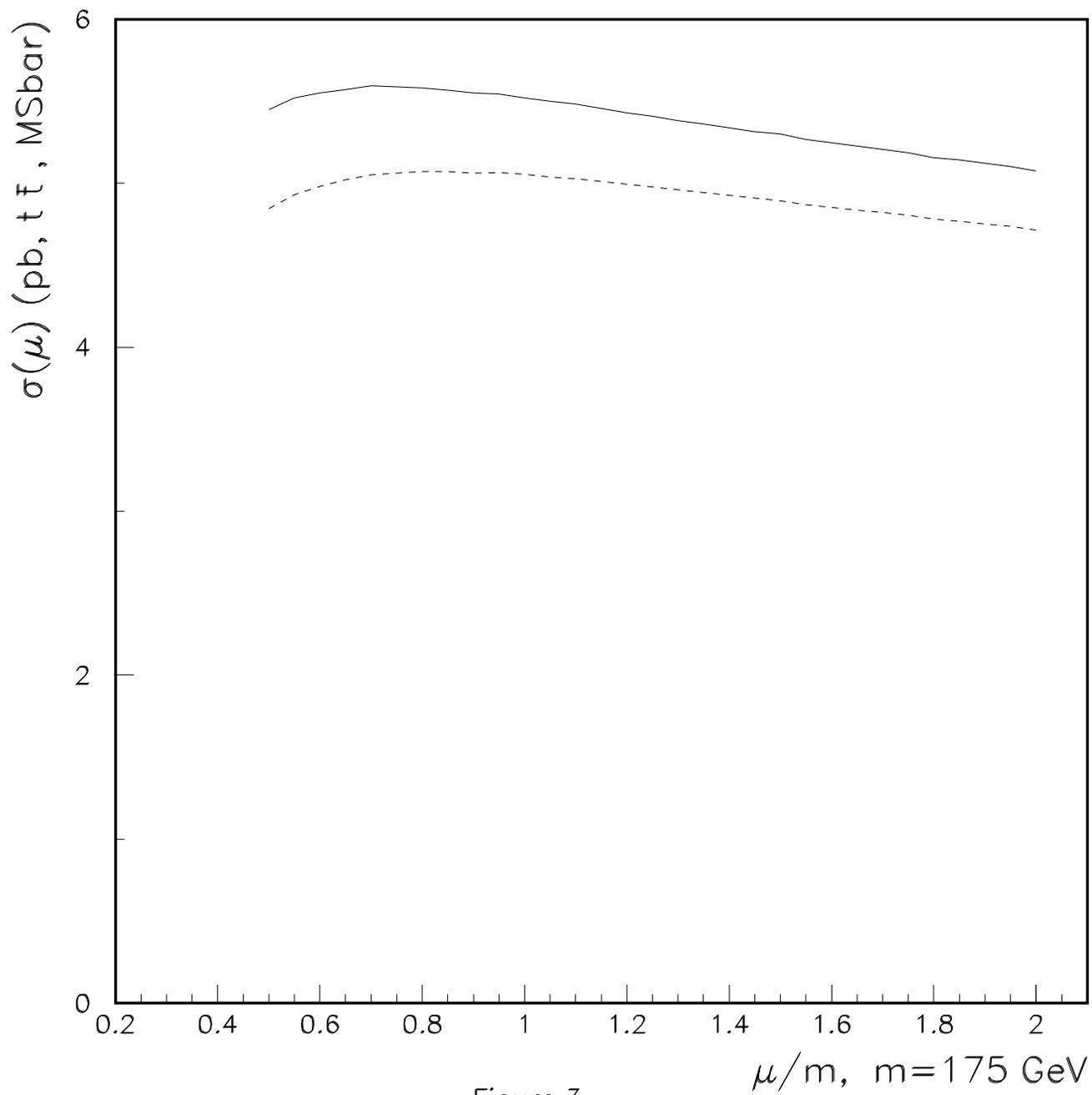

Figure 3

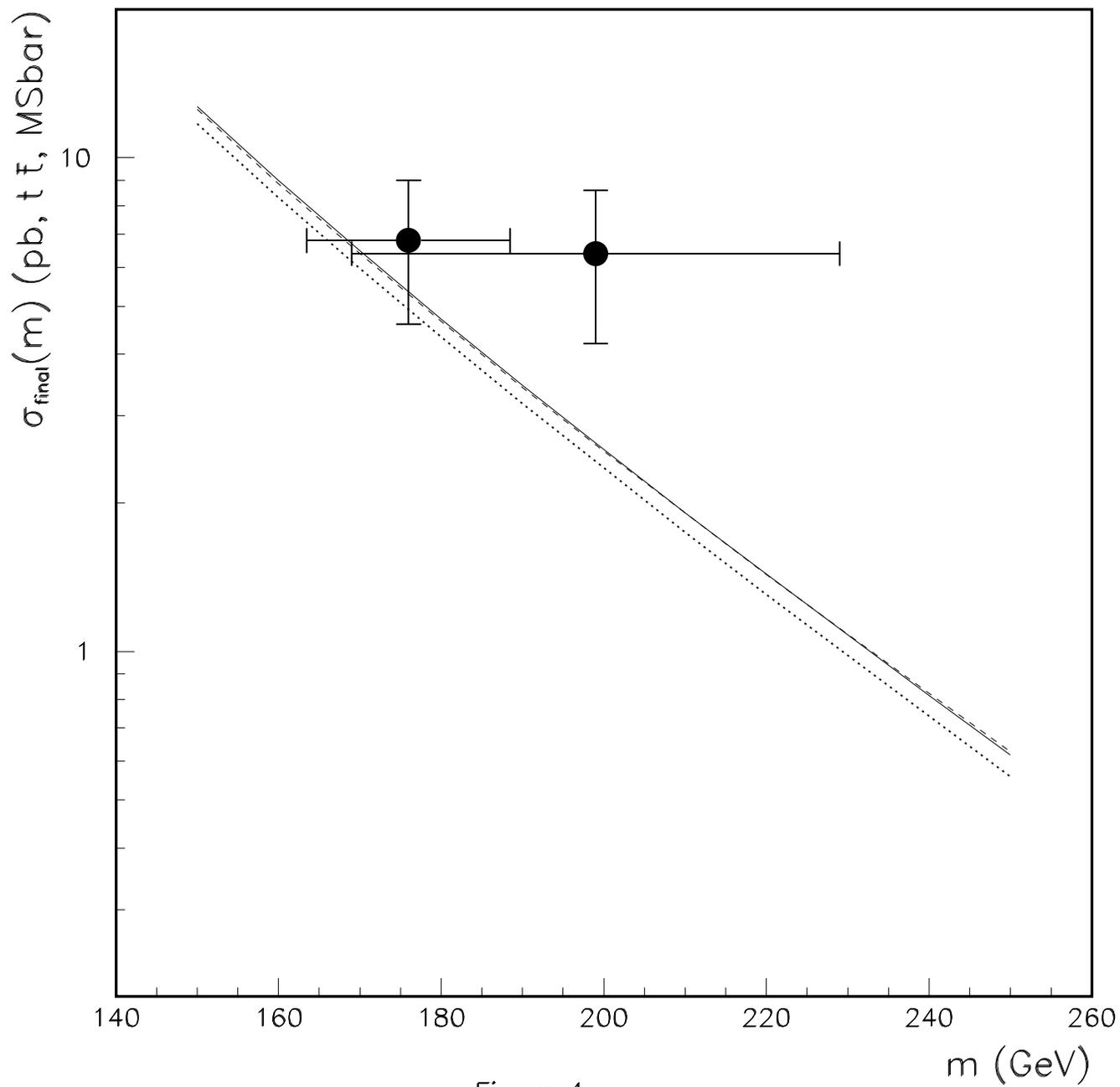

Figure 4

17